\documentclass[preprint,aps,showpacs,nofootinbib,superscriptaddress]{revtex4}
\usepackage{graphicx}
\usepackage{dcolumn}
\usepackage{amsfonts, amssymb}
\newcommand{\be}{\begin{equation}}
\newcommand{\ee}{\end{equation}}
\begin{document}

\title{Signature of the $\Lambda(1405)$ resonance in neutron spectra from the
$K^- + d $ reaction}

\author{J. R\'{e}vai\footnote{email: revai@rmki.kfki.hu}}
\affiliation{Wigner Research Center for Physics, 1525 Budapest, POB 49, Hungary}

\date{\today}

\begin{abstract}
Neutron spectra from the reaction $K^-+d\rightarrow \pi+\Sigma+n$ were calculated in the energy range $E^{cm}_{K^-}= 0-50\ MeV$ using coupled channel Faddeev equations for the description of the $\bar KNN-\pi\Sigma N$ three-body system. The aim was to trace the signature of the $\Lambda(1405)$ resonance in the spectra. We found, that while in the direct spectra kinematic effects mask completely the peak corresponding to the resonance, the deviation spectrum method \cite{AY} is able to eliminate kinematics and differentiate between different models of $\Lambda(1405)$. Four different phenomenological $\bar{K}N-\pi\Sigma$ interactions were used in order to study the effect of their pole positions on the neutron spectra.
\end{abstract}

%13.75.Jz: kaon-baryon interactions
%11.80.Jy: Faddeev equation
%25.80.Nv: kaon induced reactions and scattering
%21.45.-v: nuclear few-body systems
\maketitle

%%%%%%%%%%%%%%%%%%%%%%%%%%%%%%%%%%%%%%%%%%%%%%%%%%%%%%%%%%%%%%%%
The $\Lambda(1405)$ resonance plays a central role in low-energy kaon-nuclear physics. Being a manifestation of the assumed attraction between negative antikaons and nucleons its observability and properties are crucial for the possible existence of antikaonic nuclear clusters. Its origin and structure, mainly its one- or two-pole nature, the position and widths of these poles, are subject to vivid discussions among the representatives of different opinions, in particular, chiral perturbation theory versus phenomenology. For a review of the "state of art" see e.g. \cite{AY,Haidenbauer}  and references therein. From experimental point of view, the clarification of this problem is hindered by the fact, that this state can not be reached in two-body reactions involving stable particles, it can be observed only in processes having 3 or more particles in final states. The simplest of these is the reaction
\begin{eqnarray}
\nonumber
&{}& \nearrow K^- + d\ \textrm{or}\ K^-+p+n \\
\label{react}
K^- + d &{}& \to (\pi + \Sigma)_{I=0,1} + n\\
\nonumber
&{}& \searrow (\pi + \Sigma)_{I=1} + p
\end{eqnarray}
which has the advantage, that its dynamics can be treated exactly in the framework of the coupled particle-channels Faddeev approach. This formalism has been applied earlier for this 3-body system mainly for calculation of 3-body quasi-bound states \cite{Nina1,Sato}, or $K^-d$ scattering length \cite{Nina3,Nina4,Fayard}. The break-up reaction (\ref{react}) was studied in the early papers \cite{Toker,Torres} (of course, together with values for the $K^-d$ scattering length, too) but with the main emphasis on the possible signal of a resonance in the $\Lambda N-\Sigma N$ system, close to the $\Sigma N$ threshold. Probably at that time the  $\Lambda(1405)$ topic was not as hot as nowadays.

The only available experimental data on neutron spectra from the reaction (\ref{react}) are those of Tan \cite{Tan} from a bubble chamber experiment. Comparison with his results is shown in Fig.\ref{exp}.

The recent papers \cite{Oset1,Oset2} and \cite{Haidenbauer} are devoted to the possible observation of the $\Lambda(1405)$ in the reaction (\ref{react}). In these papers the dynamics is treated in single- plus double-scattering approximation, which might be justified  at higher incident kaon energies, but seems to be highly questionable for lower ones, as it was done in \cite{Oset2}. Our calculation is performed for low-energy kaons, in the range $E^{cm}_{K^-}=0-50\ MeV$, having in mind stopped or slowed down kaons.

The paper is organized as follows: in Sect.~II. we give a brief description of the applied formalism, Sect.~III.
contains the details and the input of the calculation, our results and their discussion are presented in Sect.~IV., while our conclusions are in Sect.~V.

%%%%%%%%%%%%%%%%%%%%%%%%%%%%%%%%%%%%%%%%%%%%%%%%%%%%%%%%%%%%%%%%%%%%%%%%%%%%%%%
\section{Formulation of the problem}
The calculation is based on the coupled-channels AGS-Faddeev treatment of the $\bar KNN-\pi\Sigma N$ three-body system. The details of this approach have been already described in detail in several papers \cite{Nina1,Sato,Nina3}, here we shall recall them only briefly, mainly to introduce the notations. The operator AGS equations for the transition operators $U_{i1}$ read
\be
\label{AGS}
U_{i1}=(1-\delta_{ij})G_0^{-1} +\sum_{j\neq i}T_jG_0U_{j1},
\ee
where $i,j=1,2,3$ are the usual pair-spectator indices,
$$
T_j=V_j+V_jG_0T_j
$$
are the two-particle $T$-operators and $G_0(z)=(z-H_0)^{-1}$ is the free Green-operator.

The configuration space $|\textbf{x}_{i}\textbf{y}_i\nu_i\rangle$ in which these operators act, apart from the usual Jacobi momentum variables $\textbf{x}_{i}\textbf{y}_i$ contain a discrete index $\nu_i=(\alpha,\sigma_i)$, which is a combination of the particle composition index $\alpha$
$$
\alpha=\{1,2,3\}=\{\bar KN_1N_2,\pi\Sigma_1N_2,\pi N_1\Sigma_2\}
$$
and an isospin label $\sigma_i=(I_iI)$ corresponding to pair isospin $I_i$ of the particle pair $i$ and total isospin $I$:
$$
\sigma_i\sim\left[\left[t_jt_k\right]^{I_i}t_i\right]^I
$$
This choice of the isospin labels corresponds to the "isospin representation", which is useful when isospin conserving pair interactions are used. Another possibility is the equivalent "charge state" or "particle" representation, characterized by the 3rd component of the particle isospins:
$$
\sigma_0\sim\{t_{1z},t_{2z},t_{3z}\}\qquad \textrm{or for $\alpha=1$}\qquad \sigma_0\sim\{\bar K^0n_1n_2,K^-p_1n_2,K^-n_1,p_2\}
$$
These representations can be transformed into each other with the help of orthogonal matrices
\be
\label{Bmat}
B^{\alpha}_{ij}=\langle\alpha\sigma_i|\alpha\sigma_j\rangle \qquad i,j=0,1,2,3
\ee
composed from 6j symbols for $i,j=1,2,3$ and from Clebsch-Gordan coefficients for ${i=0,j=1,2,3}$.
Using isospin conserving separable interactions of the form
\be
V^i=\sum_{\nu_i\nu_i'}
\delta_{I_iI_i'}\delta_{II'}|g^{I_i}_{\nu_i}\rangle\lambda^{I_i}_{\nu_i\nu_i'}\langle
g^{I_i}_{\nu_i'}|
\ee
the $T_i(z)$ operators can be written as
\be
T_i(z)=\sum_{\nu_i\nu_i'}|g^{I_i}_{\nu_i}\rangle\tau_{\nu_i\nu_i'}(z)\langle g^{I_i'}_{\nu_i'}|
\ee
with $\tau_{\nu_i\nu_i'}(z)$ being the usual (c-number) matrix, defined as:
\be
\label{tau}
(\tau_{\nu_i\nu_i'}(z))^{-1}=(\lambda^{I_i}_{\nu_i\nu_i'})^{-1}-\langle g^{I_i}_{\nu_i}|G_0(z)|g^{I_i'}_{\nu_i'}\rangle
\ee
The matrix indices $\nu_i$ in Eq.(\ref{tau}) consist of the particle space label $\alpha$ and the isospin label $\sigma_i=(I_iI)$. Due to isospin conservation of our interactions, the coupling constant matrix $\lambda$ is diagonal in $(I_iI)$, while for particle pairs $i$ capable to change their identity $(\bar KN\leftrightarrow\pi\Sigma)$ it has non-diagonal elements in the particle labels $(\alpha\alpha')$. As for the matrix elements of $G_0$, it does not change particle identities, thus it is diagonal in $\alpha$, and if we take averaged masses for particles within an isospin multiplet, it is also diagonal in pair- and total isospin indices $(I_iI)$. However, if physical (unequal) masses are used,
$G_0$ will be diagonal only in "particle" representation, while in the $\sigma_i$ "isospin" representation it will acquire non-diagonal elements both in $I_i$ and $I$, proportional to the mass differences.

The equations (\ref{AGS}) for the transition operators take the form
\be
U_{i1}=(1-\delta_{ij})G_0^{-1} +\sum_{j\ne i}\sum_{\nu_j'}|g_{\nu_j}\rangle \tau_{\nu_j,\nu_j'}\langle X^j_{\nu_j'}|\qquad \textrm{with}
\qquad \langle X^j_{\nu_j'}|=\langle g^{I_j}_{\nu_j'}|G_0U_{j1}
\ee
Introducing the functions $X^j_{\nu_j}(\textbf{y}_j)=\langle X^j_{\nu_j}|\Phi_0\rangle$, where $\textbf{y}_i$ is the momentum of the spectator particle, corresponding to the pair $j$ and $|\Phi_0\rangle=|\varphi_dP_{K}\rangle$ is the initial state with the deuteron wave function $|\varphi_d\rangle$ and $P_K$ - the momentum of the incident kaon, we get the set of integral equations:
\be
\label{system}
X^i_{\nu_i}(\textbf{y}_j)=(1-\delta_{i1})\langle g^{I_i}_{\nu_i}|
\Phi_0\rangle+\sum_{j\ne i}\sum_{\nu_j,\nu_j'}\int Z_{\nu_i\nu_j}(\textbf{y}_i
,\textbf{y}_j)\tau_{\nu_j\nu_j'}(z-y^2_j/2\mu_{j,ki})X^j_{\nu_j'}(\textbf{y}_j)d\textbf{y}_j
\ee
with the kernel
\be
Z_{\nu_i\nu_j}(\textbf{y}_i,\textbf{y}_j)=\langle g^{I_i}_{\nu_i}|G_0(z)|g^{I_j}_{\nu_j}\rangle .
\ee
The size of the system (\ref{system}) can be reduced by introducing symmetric (antisymmetric) combinations of X-functions, with respect to interchange of baryon numbering. The baryon spins do not enter explicitly in this formalism, therefore the total baryon spin $S$ remains unchanged in the process (is a conserved quantum number). For a given $S$ value the total antisymmetry required by the Pauli principle has to be ensured by the space-isospin part.  Thus for $S=0$ ($K^-pp$ system) we have to work with the symmetric combinations of $X$-s, while for $S=1$ (our $K^-d$ system) the antisymmetric combinations are needed. As a result, the labeling of the unknown functions $\nu_i=(\alpha\sigma_i)$ is changed to ${\mu_a=(a,\sigma_a)}$, where $a$ denotes a pair of interacting particles, irrespectively to which original particle composition channel they belonged and $\sigma_a$ denotes the corresponding isospin values. Thus we are left with $X_{a\sigma_a}(\textbf{y}_a)$ and $a$ can take the values $\bar KN,NN,\Sigma N$ and $\pi\Sigma$ ($\pi N$ is missing, since we neglected the $\pi N$ interaction, see next section).

The break-up transition operator $U_{01}$ can be expressed in terms of the $U_{i1}$-s as:
$$
U_{01}={1\over 2}(U_{11}+U_{21}+U_{31})
$$
and the break-up amplitude reads
\be
\label{BU}
A_{BU}=\langle\Phi_f|U_{01}|\Phi_0\rangle .
\ee
For the reaction under consideration the properly antisymmetrized final state is
$$
|\Phi_f\rangle=|\textbf{x}_{\pi\Sigma},\textbf{y}_N;\sigma_{\pi\Sigma}\rangle=
{1\over\sqrt{2}}(|\textbf{x}_{\pi\Sigma_1},\textbf{y}_{N_2};\sigma_{\pi\Sigma_1}\rangle-
|\textbf{x}_{\pi\Sigma_2},\textbf{y}_{N_1};\sigma_{\pi\Sigma_2}\rangle)
$$
The break-up amplitude can be expressed in terms of the X-functions as
\begin{eqnarray}
\label{ABU}
&A_{BU}(\textbf{x}_{\pi\Sigma},\textbf{y}_N;\sigma_{\pi\Sigma})
 =
\nonumber &\\
\label{BUamp}
&-g_{\pi\Sigma}(\textbf{x}_{\pi\Sigma})
\left[\tau_{\pi\Sigma,\bar KN}(z-y^2_N/2\mu_{N,\pi\Sigma})X_{\bar KN}(\textbf{y}_N)+
\tau_{\pi\Sigma,\pi\Sigma}(z-y^2_N/2\mu_{N,\pi\Sigma})X_{\pi\Sigma}(\textbf{y}_N)\right] &\\
&-B^2_{31}g_{\Sigma N}(u\textbf{y}_N+v\textbf{x}_{\pi\Sigma})\tau_{\Sigma N,\Sigma N}
(z-|\textbf{x}_{\pi\Sigma}-w\textbf{y}_N|^2/2\mu_{\pi,\Sigma N})X_{\Sigma N}
(\textbf{x}_{\pi\Sigma}-w\textbf{y}_N)\nonumber &,\
\end{eqnarray}
where $B^2_{31}$ is an isospin recoupling matrix (see Eq.(\ref{Bmat})), $u,v\ \textrm{and}\ w$ are mass coefficients of the transformation between Jacobi momentum sets. In Eq.(\ref{BUamp}) we omitted the isospin labels, the quantities are vectors (matrices) in isospin space. The on-shell amplitude for a given neutron energy $E_n$ depends on $E_n,t$ and the isospin labels $\sigma_{\pi\Sigma}$:%\goodbreak
$$
A(E_n,t,\sigma_{\pi\Sigma})=A_{BU}(\textbf{x}_{\pi\Sigma},\textbf{y}_N;\sigma_{\pi\Sigma})
$$
with
$$
|\textbf{y}_n|=\sqrt{2E_n\mu_{N,\pi\Sigma}};\
|\textbf{x}_{\pi\Sigma}|=\sqrt{2(E_{\pi\Sigma N}-E_n)\mu_{\pi\Sigma}};\
t=\cos(\textbf{x}_{\pi\Sigma},\textbf{y}_N)
$$
%\begin{table}[t]
%begin{tabular}{c|l}
%$A(E_n,t,\sigma_{\pi\Sigma})=A_{BU}(\textbf{x}_{\pi\Sigma},\textbf{y}_N;\sigma_{\pi\Sigma})\ $ & $\ %|\textbf{y}_n|=\sqrt{2E_n\mu_{N,\pi\Sigma}}$\\
% & $\ |\textbf{x}_{\pi\Sigma}|=\sqrt{2(E_{\pi\Sigma N}-E_n)\mu_{\pi\Sigma}}$\\
% & $\ t=\cos(\textbf{x}_{\pi\Sigma},\textbf{y}_n)$\\
%\end{tabular}
%\end{table}

The physically observable final state corresponds to a certain particle composition, not to a definite isospin state, therefore the amplitude has to be transformed into the $\sigma_0$ representation, using the suitable $B$ matrix of Eq.(\ref{Bmat}):
{$$
A(E_n,t,\sigma_0)=\sum_{\sigma_{\pi\Sigma}}\left(B^2_{03}\right)_{\sigma_0,\sigma_{\pi\Sigma}}A(E_n,t,\sigma_{\pi\Sigma}),
$$
where $\sigma_0$ can be $\{\pi^+\Sigma^-n,\pi^0\Sigma^0n,\pi^-\Sigma^+n\}$.} The neutron spectrum is proportional to the differential cross section
\be
\label{cross}
P(E_n,t,\sigma_0)\sim{d\sigma\over d\Omega_{\textbf{x}_{\pi\Sigma}}d\Omega_{\textbf{y}_N}dE_n}=
(2\pi)^4\mu_{\pi\Sigma}\mu_{N,\pi\Sigma}\mu_{K,NN}{x_{\pi\Sigma}y_N\over P_K}|A(E_n,t,\sigma_0)|^2
\ee
The inclusive neutron spectrum (when no other particles are detected) is given by
\be
\label{PEn}
P(E_n)=\sum_{\sigma_0}\int_{-1}^{1}dtP(E_n,t,\sigma_0)
\ee
The above considerations refer to the neutrons emerging from the reaction $K^-+d\rightarrow \pi+\Sigma+n$; when the energy of the incident kaon exceeds the deuteron binding energy, neutrons are also emitted from the reaction  $K^-+d\rightarrow K^-+p+n$. Their spectrum can be deduced in a similar way to Eq.(\ref{BUamp}) from the $X_{\bar KN}$ and $X_{NN}$ functions. The allowed energy range for neutrons from the first reaction is $(0,E_K+E_d+\Delta)$, while for the second it is $(0,E_K+E_d)$, where $\Delta$ is the difference of the $\bar KN$ and $\pi\Sigma$ threshold energies.

%%%%%%%%%%%%%%%%%%%%%
\section{Details of the calculation and the input}
The main purpose of the present work is to study the possible signature(s) of the $\Lambda(1405)$ resonance in the neutron spectra from the reaction (\ref{react}). In our calculation we used the two-body interactions of \cite{Nina3} and \cite{Nina4}, which are adjusted for our three-body model. They are $s$-wave, separable isospin dependent and isospin conserving interactions with Yamaguchi type form-factors.

In particular, for the two-channel $\bar KN-\pi\Sigma$ interaction we used two variants, both having a one and a two pole version for the  $\Lambda(1405)$.  They both reproduce all available experimental data on the low-energy $\bar KN$ system, the first one is fitted to the KEK data on the kaonic hydrogen $1s$ level shift, while the second one reproduces the most recent SIDDHARTA data. Their pole positions  are shown in Table \ref{poles}.

\begin{table}[h]
\caption{\label{poles}Pole positions of the $\bar{K}N-\pi\Sigma$ potentials (in MeV), the negative real parts correspond to distances from the $\bar{K}N$ threshold.}
%\begin{ruledtabular}
\begin{center}
\begin{tabular}{c|c|c}
\hline\hline
& KEK\ & SIDDHARTA\\
\hline
1-pole\ \ &$-23.6-35.6$\ \textbf{i}\ ($1411.0-35.6$\  \textbf{i})&$-6.4-46.8$\ \textbf{i}\ ($1428.1-46.8$\  \textbf{i})\\
2-pole\ \ &$-22.2-36.3$\ \textbf{i}\ ($1412.4-36.3$\  \textbf{i})&$-14.8-57.2$\ \textbf{i}\ ($1419.8-57.2$\  \textbf{i})\\
{}&$-58.8-102.5$\ \textbf{i}\ ($1375.8-102.5$\  \textbf{i})&$-56.6-101.8$\ \textbf{i}\ ($1380.0-101.8$\  \textbf{i})\\
\hline
\end{tabular}
\end{center}
%\end{ruledtabular}
\end{table}

The numbers in Table \ref{poles} differ slightly from those given in the original papers \cite{Nina3,Nina4}. The reason is, that the above ones were calculated with averaged masses and without Coulomb interaction - as they appear in most of the 3-body calculations, - while the fitting to the experimental data was performed with physical masses and Coulomb interaction. Since the main aim of the present work is to study the appearance of subthreshold resonances of different type in a 3-body reaction, we kept both interactions, not only the more advanced one.

The triplet $NN$ interaction is a two-term one to account for the short range repulsion, with form-factors fitted to reproduce the deuteron binding energy and $s$-wave phase shifts.
The $S=1$ $\Sigma N$ interaction in the $I=1/2$ isospin state is complex, since it was deduced from a two-channel $\Sigma N-\Lambda N$ interaction, while for $I=3/2$ it is real.
The $\pi N$ interaction was neglected in our calculation due to its weak $s$-wave part.

The total angular momentum was restricted to $L=0$ since we believe, that for our $s$-wave interactions the essential dynamics can be traced in spite of this limitation. Keeping the interactions $s$-wave, the extension to higher angular momenta is straightforward, unlike the case of inclusion of $p,d,..$-wave interactions.

We considered incident kaon energies in the interval $E^{cm}_{K^-}=0-50\  MeV$ ${(P^{LAB}_K \sim 0-250\ MeV/c)}$. The system of integral equations (\ref{system}) in the case of physical masses ($I=1/2$ and $I=3/2$ mixed) consists of 12 equations, while for averaged masses - of 8.

As a numerical method we used expansion of the unknown functions on a cubic spline basis, for the distribution of nodes and collocation points the prescription of \cite{Svenne} was used with a slight modification to allow nonsymmetric intervals and distributions on the two sides of the break-up singularity. Complete convergence of the results was achieved for $\sim 20$ nodes in the non-break-up channels, while for the break-up channels $\sim 30-35$ nodes were necessary. Apart from the lower dimensionality of the matrices to be inverted, the use of spline expansion is especially advantageous when break-up amplitudes are calculated, since no interpolation of the solutions is needed.

%%%%%%%%%%%%%%%%%%%%%%%%%%%%%%%%%%
\section{Results and discussion}
We start the presentation of our results by a "by-product": the effect of the physical versus averaged masses of the $\bar K^0$ and $K^-$ mesons on the $K^-d$ scattering length. The inclusion of the possibility of isospin mixing due to this mass difference allowed us to extend the results of \cite{Nina3,Nina4} in this respect. Our results for the two potential versions are shown in Table \ref{scatlen}. The results for averaged masses coincide with those of \cite{Nina3,Nina4}, while for physical masses they differ by a few per cent, mainly in the real part. At present level of accuracy of available information - both theoretical and experimental - on the $\bar KN$ interaction and $\bar K$-nuclear clusters this difference does not seem to be essential. However, once it might become useful to have some numerically reliable information on the order of magnitude of this effect.
\begin{table}[h]
\caption{\label{scatlen}$K^-d$ scattering lengths of the $\bar{K}N-\pi\Sigma$ potentials for physical and averaged masses of $K^-$ and $\bar{K}^0$ (in fm).}
%\begin{ruledtabular}
\begin{center}
\begin{tabular}{c|c|c|c|c}
\hline
\hline
&\multicolumn{2}{c|}{KEK}&\multicolumn{2}{c}{SIDDHARTA}\\\cline{2-5}
&\ averaged\ &\ physical\ &\ averaged\ &\ physical\ \\
\hline
1-pole\ \ &$-1.49+0.97$\ \textbf{i} &$-1.52+0.98$\ \textbf{i}&$-1.47+1.22$\ \textbf{i} &$-1.50+1.23$\ \textbf{i}\\
2-pole\ \ &$-1.57+1.10$\ \textbf{i} &$-1.60+1.12$\ \textbf{i}&$-1.50+1.23$\ \textbf{i} &$-1.54+1.24$\ \textbf{i}\\
\hline
\end{tabular}
\end{center}
%\end{ruledtabular}
\end{table}

On the other hand, when calculating neutron spectra our interest was focused on qualitative signals of the $\Lambda(1405)$ in the line shapes, therefore we used averaged masses, what simplified the numerical work to some extent.

We have calculated the inclusive neutron spectra $P(E_n)$ (\ref{PEn}) for different incident kaon energies, both below and above the deuteron break-up threshold.

\begin{figure}[h]
\centering
\includegraphics[width=\textwidth]{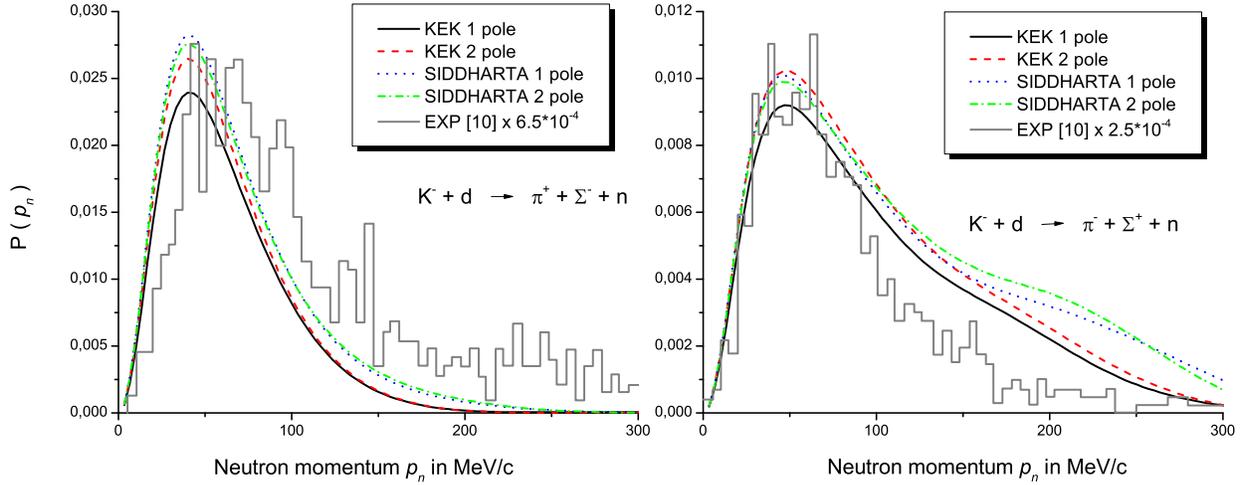}
\caption{
Momentum distribution of the neutrons from the reaction $K^-+d\rightarrow\pi^\pm +\Sigma^\mp + n$ for stopped kaons
}
\label{exp}
\end{figure}

In order to allow a comparison with the only available experimental data in our energy range we performed a calculation for $E_K=0$ (this was also needed for the scattering lengths). The results are shown in Fig.\ref{exp}.
Unfortunately, this can be called only a "quasi-comparison", since it is not clear, what calculated theoretical quantity should be compared with the data shown on their Fig.2. If the reaction is "at rest", then probably it starts from an atomic orbit, which case necessitates a somewhat different Faddeev treatment. Our calculation can imitate the "at rest" criterion by taking $E_K=0$, but then our cross section formula (\ref{cross}) must be modified: the incoming flux normalization (division by $P_K$) has to be removed and an extra $y_N$factor must be added to get momentum distribution instead of energy spectrum. Still we believe, that the curves displayed on Fig.\ref{exp}. qualitatively correspond to the same momentum distribution, but have different absolute normalization. Therefore, to bring the curves together, the experimental ones were scaled down, as indicated in the captions. The difference of the scaling factors (and of the arbitrary units on the y-axis) corresponds to the fact, that the number of neutrons coming with $\Sigma^-$ exceeds the number of those, emitted with $\Sigma^+$ by a factor of $\sim 2.5$ (in \cite{Tan} it was estimated as $2$). The agreement can be considered as acceptable, especially having in mind the experimental uncertainties. However, due to the practical indistinguishability of the theoretical curves, from the point of view of the signature of the $\Lambda(1405)$ in this reaction, this agreement seems to be of not much help.

\begin{figure}[h]
\centering
\includegraphics[width=\textwidth]{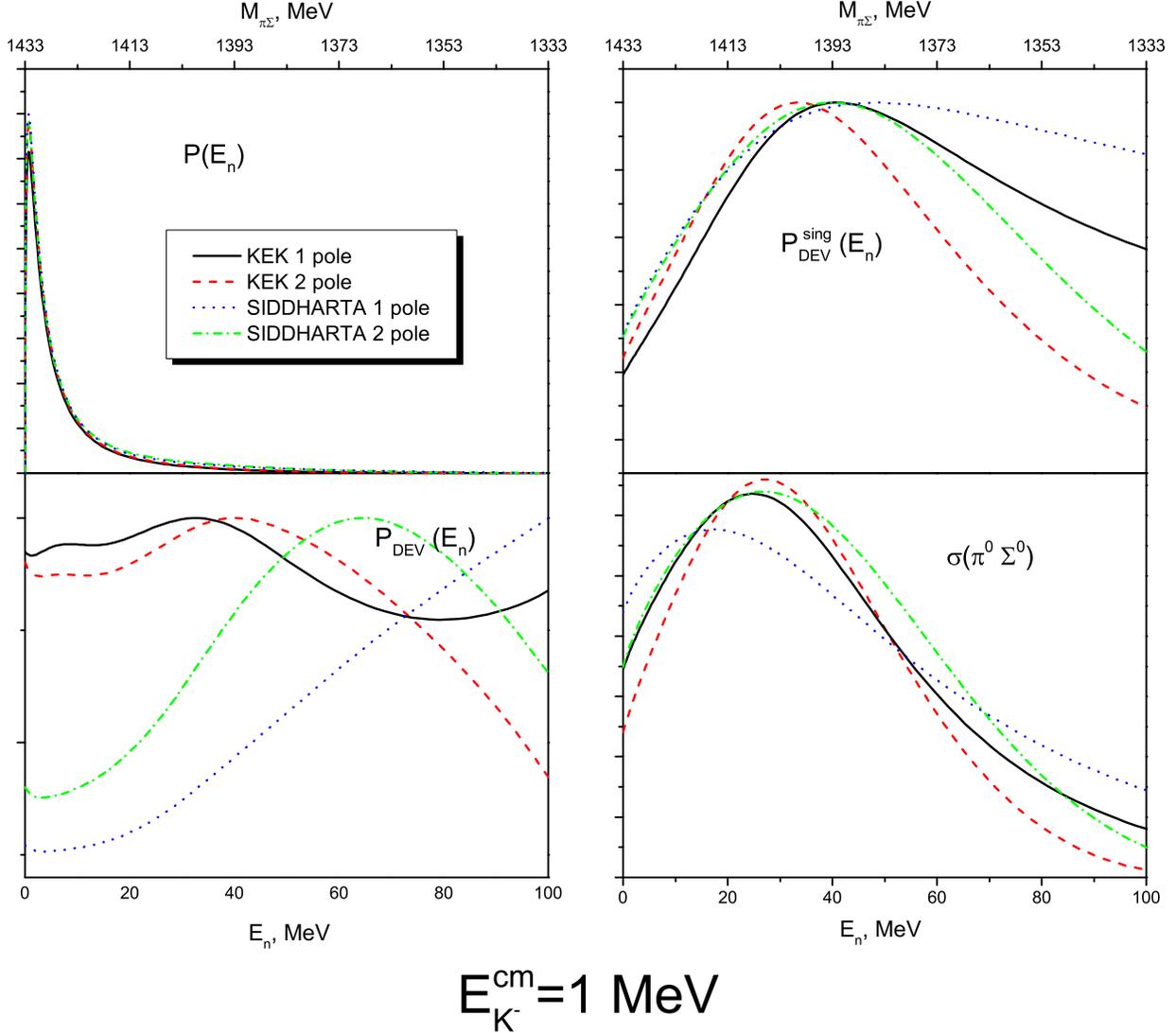}
\caption{Neutron spectra for kaon incident energy $E^{cm}_{K^-}=1\ MeV$.
Upper left graph - direct neutron spectra $P(E_n)$, eq.(\ref{PEn});
lower left graph - deviation spectrum $P_{DEV}(E_n)$, eq.(\ref{DEV1});
upper right graph - single scattering deviation spectrum $P_{DEV}^{sing}(E_n)$, eq. (\ref{DEV1});
lower right graph - "original" $\Lambda(1405)$ as $\pi^0\Sigma^0$ elastic cross section.
}
\label{fig1}
\end{figure}

\begin{figure}[h]
\centering
\includegraphics[width=\textwidth]{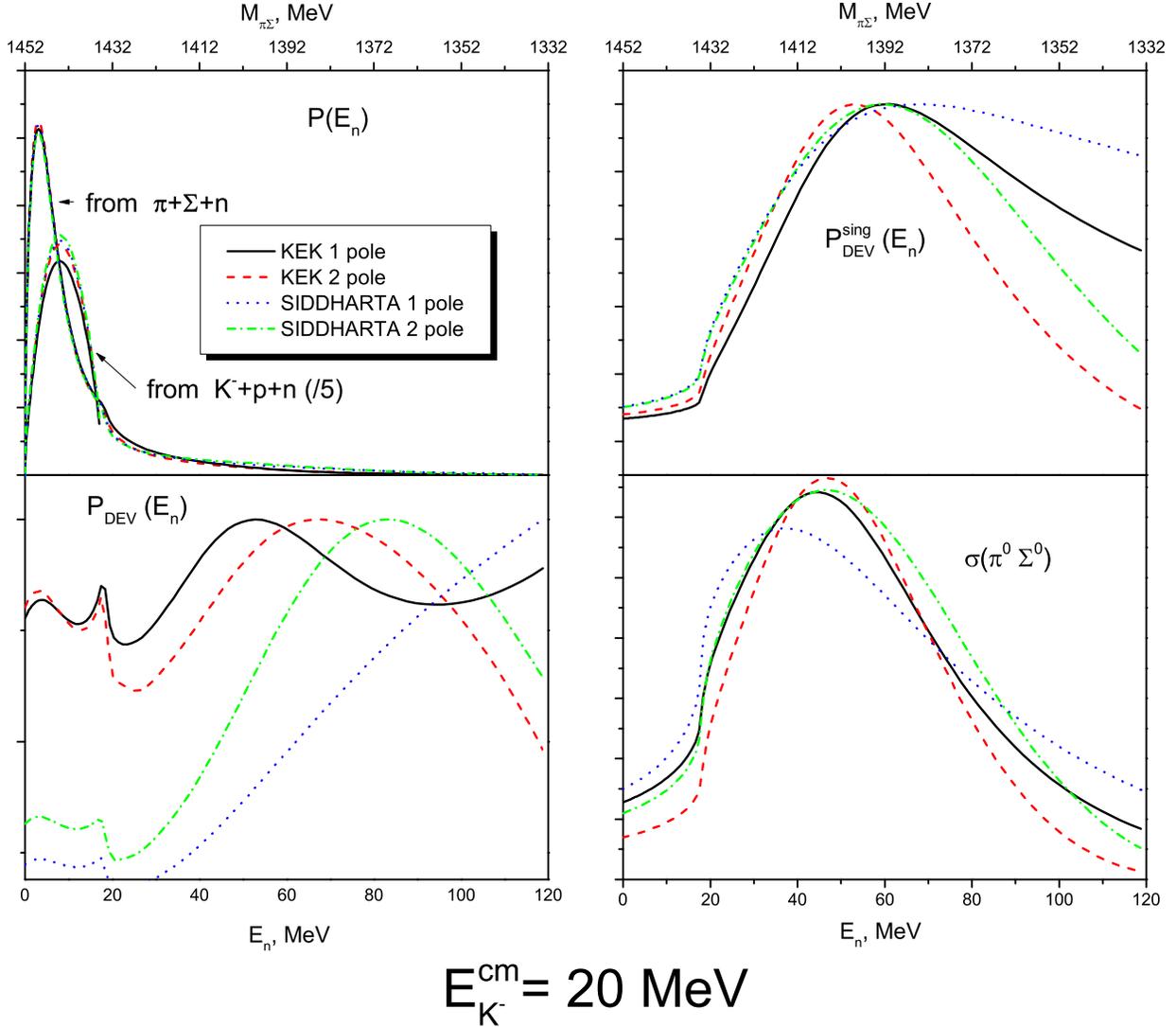}
\caption{Neutron spectra for kaon incident energies $E^{cm}_{K^-}=20\ MeV$ The four graphs are the same as in Fig.\ref{fig1}.
\label{fig2}}
\end{figure}

\begin{figure}[h]
\centering
\includegraphics[width=\textwidth]{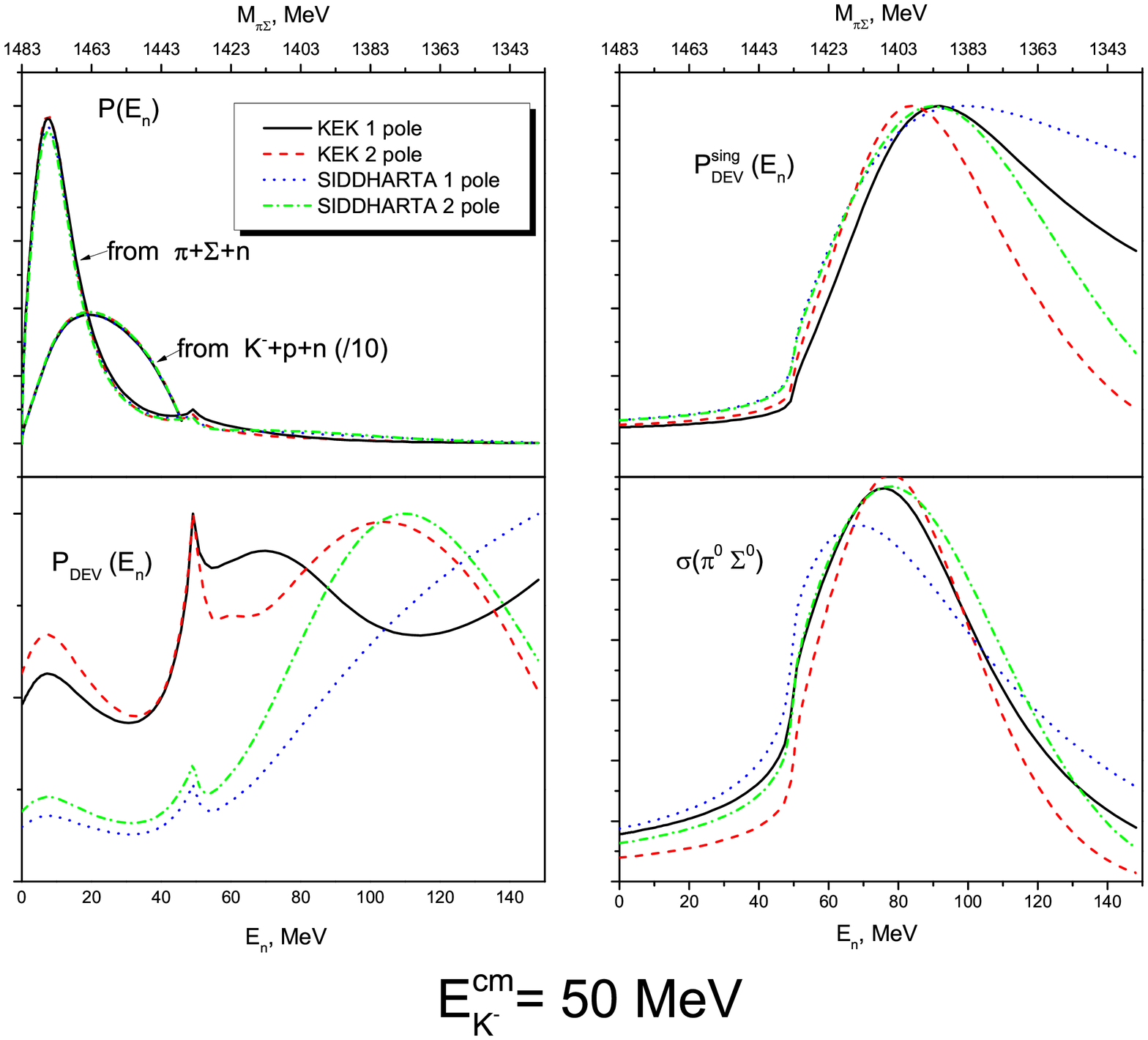}
\caption{Neutron spectra for kaon incident energies $E^{cm}_{K^-}=50\ MeV$ The four graphs are the same as in Fig.\ref{fig1}.
\label{fig3}}
\end{figure}

For kaon incident energies $E_{K^-}^{cm}=1,20,50 MeV$ the results are displayed in Figs. \ref{fig1}, \ref{fig2} and \ref{fig3}, respectively, (upper left graphs)\footnote{Since our main concern is the possible trace of the $\Lambda(1405)$ in the line shapes of the calculated spectra, the arbitrary units on the $Y$-axes of our graphs are chosen to optimize visibility.}.

%%%%%%%%%%%%%%%%%%%%%%%%%%%%%%%%%%%%%%%%%%%%%%%%%%%%%%%%%%%%%%%%%%%%%%%%%%
%%%%%%%%%%%%%%%%%%%%%%%%%%%%%%%%%%%%%%%%%%%%%%%%%%%%%%%%%%%%%%%%%%%%%%%%%%

The overall shape of the spectra is a strong peak near the origin with no signal of the $\Lambda(1405)$ resonance. The direct $P(E_n)$ spectra are practically indistinguishable for the four considered $\bar{K}N-\pi\Sigma$ potentials. For kaon energies above the deuteron binding energy there are two modifications: the neutron spectra from the
$K^-+d\to \pi+\Sigma+n$ channel show a cusp at neutron energies $E_n=E_{th}$ when the $\bar KN$ system is at its threshold, and additional neutrons show up from the $K^-+d\to K^-+p+n$ reaction in a form of a structureless
%%%%%%%%%%%%%%%%%%%%%%%%%%%%%%%%%%%%%%%%%%%%%%%%%%%%%%%%%%%%%%%%%%%%%%%%%%%%
%%%%%%%%%%%%%%%%%%%%%%%%%%%%%%%%%%%%%%%%%%%%%%%%%%%%%%%%%%%%%%%%%%%%%%%%%%%%%%%%%
bump between $E_n=0$ and $E_n=E_{th}$ (on the graphs it is scaled down to allow to draw it on the same plot with the other neutrons). The reason, why the $\Lambda(1405)$  is not seen in these spectra is essentially kinematical:
the neutron energy in the resonance region should exceed the incident energy of the kaon  by the amount of energy, which separates the pole position from the $\bar KN$ threshold, while in the deuteron the neutron energy (momentum) distribution is dominated by the low energy part.

%%%%%%%%%%%%%%%%%%%%%%%%%%%%%%%%%%%%%%%%%%%%%%%%%%%%%%%%%%%%%%%%%%%%%%%%%
%%%%%%%%%%%%%%%%%%%%%%%%%%%%%%%%%%%%%%%%%%%%%%%%%%%%%%%%%%%%%%%%%%%%%%%%%%%%

In order to eliminate this kinematical "inconvenience" Esmaili,  Akaishi and Yamazaki (EAY) \cite{AY} propose to consider instead of $P(E_n)$ the deviation spectrum:
\be
\label{DEV}
P_{DEV}={P(E_n)\over P_{nonres}(E_n)}
\ee
In Eq.(\ref{DEV}) $P_{nonres}(E_n)$ is a non-resonant background spectrum, containing the kinematics of the reaction. Let's see, how this idea can be realized in our case.

Considering the zero-order iteration of our (symmetrized) system of integral equations (\ref{system}), the only non-vanishing $X$ will be the inhomogeneous term:
\be
\label{sing}
X_{\bar {K}N}(y_N)=\langle g_{\bar {K}N}|\Phi_0\rangle=\langle g_{\bar {K}N}|\varphi_d P_K\rangle .
\ee
This ansatz is usually called single scattering approximation. Substituting eq.(\ref{sing}) into eq.(\ref{ABU}) we get the corresponding break-up amplitude:
\begin{eqnarray}
\label{ABUsing}
A_{BU}^{sing}(x_{\pi\Sigma},y_N;\sigma_{\pi\Sigma})
&=&g(x_{\pi\Sigma})\tau_{\pi\Sigma,\bar{K}N}(z-y_N^2/2\mu_{N,\pi\Sigma})\langle g_{\bar {K}N}|\varphi_d
P_K\rangle\nonumber\\
&=&\langle x_{\pi\Sigma},y_N|T_{\pi\Sigma,\bar{K}N}|\varphi_d P_K\rangle,
\end{eqnarray}
which is the matrix element of the two-body $T$ operator between the initial and final state. It contains two-body dynamics through the $T$ operator and the kinematical input: the transformation between the Jacobi-coordinates and the deuteron wave function. This is basically the formula, which EAY used to calculate the transition amplitude from the $K^-d$ atomic state to the $\pi\Sigma n$ continuum. As for the non-resonant amplitude they suggest to replace $T_{\pi\Sigma,\bar{K}N}$ in (\ref{ABUsing}) by $V_{\pi\Sigma,\bar{K}N}$:
\be
\label{born1}
A_{BU}^{Born}(x_{\pi\Sigma},y_N;\sigma_{\pi\Sigma})=\langle x_{\pi\Sigma},y_N|V_{\pi\Sigma,\bar{K}N}|\varphi_d P_K\rangle,
\ee
that is, to use the Born-approximation, which contains all the kinematics.

Thus we have three amplitudes with the properties
\begin{itemize}
\item[-]{$A_{BU}\ \longrightarrow$ three-body dynamics + three-body kinematics,}
\item[-]{$A_{BU}^{sing}\ \longrightarrow$ two-body dynamics + three-body kinematics,}
\item[-]{$A_{BU}^{Born}\ \longrightarrow$ \qquad\qquad\qquad\qquad\qquad three-body kinematics,}
\end{itemize}
and we expect, that the DEV spectra
\be
\label{DEV1}
P_{DEV}(E_n)=P(E_n)/P^{Born}(E_n)\ ;\ \ P_{DEV}^{sing}(E_n)=P^{sing}(E_n)/P^{Born}(E_n)
\ee
will display (reveal) three- and two-body dynamics, respectively.

\begin{figure}[h]
\centering
\includegraphics[width=\textwidth]{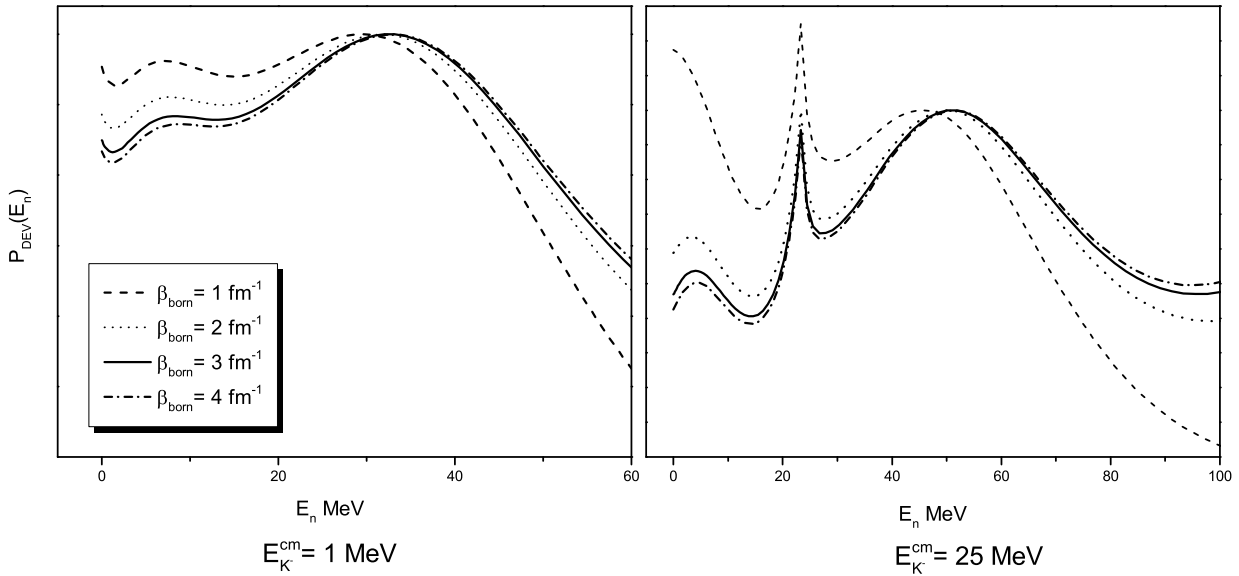}
\caption{Effect of $\beta_{Born}$ on the deviation spectra
\label{born}}
\end{figure}

 It is assumed, that the $E_n$ dependence of $P_{nonres}(E_n)=P^{Born}(E_n)$ is basically determined by the features of the initial and final states, while the details of the $V_{\bar KN,\pi\Sigma}$ potential (within reasonable limits) influence it only weakly. This expectation is important, if the deviation spectrum method is to be applied for extracting some information on $\Lambda(1405)$ from an experimentally measured neutron spectrum. (Hopefully, the matrix element (\ref{born1}) can be calculated in an experimental group, too.) To check this anticipated model-independence of the method, we calculated  $P^{Born}(E_n)$ \emph{not} with our realistic $\bar KN-\pi\Sigma$ interactions, but with the simplest possible separable potential:
$$
\langle x_{\pi\Sigma}|V^I_{Born}|x_{\bar KN}\rangle={1\over x_{\pi\Sigma}^2+(\beta_{\pi\Sigma}^I)^2}\lambda_{\pi\Sigma,\bar KN}^I{1\over x_{\bar KN}^2+(\beta_{\bar KN}^I)^2}
$$
and took
$$
\lambda_{\pi\Sigma,\bar KN}^{I=0}=\lambda_{\pi\Sigma,\bar KN}^{I=1}=1\ , \beta_{\pi\Sigma}^{I=0}=\beta_{\pi\Sigma}^{I=1}=\beta_{\bar KN}^{I=0}=\beta_{\bar KN}^{I=1}=\beta_{Born}
$$
We calculated the $P_{DEV}(E_n)$ deviation spectra for different $\beta_{Born}$ values and incident kaon energies. Typical results are shown in Fig. \ref{born}.
%%%%%%%%%%%%%%%%%%%%%%%%%%%%%%%%%%%%%%%%%%%%%%%%%%%
%%%%%%%%%%%%%%%%%%%%%%%%%%%%%%%%%%%%%%%%%%%%%%%%%%%%%%%%%%%%%%%%%%%%%%%%%%
The shape of the $P_{DEV}$ does not depend on the normalization of $P_{nonres}$ (see the arbitrary choice of the $\lambda_{\pi\Sigma,\bar KN}$-s) and we have normalized the $P_{DEV}$ curves to have their maxima at 1. It can be seen, that for reasonable $\pi\Sigma-\bar KN$ interaction range values the spectra show practically no dependence on $\beta_{Born}$, thus confirming the presumed (approximate) model-independence of the  $P_{DEV}$ method. According to the foresaid, the rest of our calculations were done with $\beta_{Born}=3 fm^{-1}$.

Figs. 1.,2. and 3. demonstrate our main results. Apart from the direct spectra $P(E_n)$, which show no trace of the $\Lambda(1405)$, we show also the full- and single-scattering deviation spectra, too. For comparison, we plotted also the (hypothetical) $\pi^0\Sigma^0$ elastic cross sections as calculated from the $\bar{K}N -\pi\Sigma$ interaction for the corresponding $\pi\Sigma$ energy (top scale). Since this cross section "feels" only the $I=0$ part of the $\bar{K}N -\pi\Sigma$ interaction, its peak is usually identified with the $\Lambda(1405)$ resonance. Thus the similarity of the $P_{DEV}(E_n)$ and $\sigma(\pi^0\Sigma^0)$ line shapes can tell us about the reliability of extracting information about $\Lambda(1405)$ from the reaction under consideration. Obviously, the $P_{DEV}^{sing}$ spectra show more similarity with the original $\Lambda(1405)$ shape than the full $P_{DEV}(E_n)$ spectra, since their dynamical content is more or less the same. For three of the four considered potentials (KEK 1, KEK 2, SIDDHARTA 2) a clear resonant structure can be seen in the full deviation spectra, however, the shapes and positions can significantly differ from their "originals".

As for the fourth potential, SIDDHARTA 1, its deviation spectra do not show any signature of its original $\Lambda(1405)$, although there are clean maxima in the corresponding $P_{DEV}^{sing}$ and $\sigma(\pi^0\Sigma^0)$ curves. The reason might be the extreme closeness of the pole to the $\bar{K}N$ threshold combined with its large width.

How can we interpret these results?

As for the direct spectra, we can say, that they are practically indistinguishable for all four potentials, having quite different
pole structure, - at least in the considered energy range - and thus are useless for differentiating between models of $\Lambda(1405)$.

The deviation spectra in single scattering approximation show a remarkable similarity with the "original" $\sigma(\pi^0\Sigma^0)$ cross section
curve. Since the authors of \cite{AY} used this approximation for the solution of their three-body problem, this similarity lead them
to optimistic conclusions about the general and simple applicability of the DEV spectrum method.

Unfortunately, in the case of DEV spectra calculated with the true three-body operators, this similarity does not hold any more.
The most remarkable observation, concerning these spectra, however, is, that they are quite sensitive to the choice of the $\bar{K}N$
potentials, or to their pole structure. This in principle allows to distinguish between potentials leading to different pole positions.
However, this possibility does not mean, that visual observation of maxima in the DEV spectra can be used for identification of the $\Lambda(1405)$ pole positions. Even in the two-body case the connection between cross section maxima and pole positions are far from being trivial, except for very narrow resonances. This can be seen e.g. from comparison of the single-scattering DEV spectra and the $\sigma(\pi^0\Sigma^0)$ cross
sections, which are similar, proving, that they describe essentially the same resonance, however, their maxima and widths are not simply
related to the pole parameters as they are shown in Table \ref{poles}.

In the three-body case the observed
spectra can be related to the two-body characteristics of the input potentials only via reliable dynamic calculations, which in the low-energy regime mean solution of Faddeev equations. Their results, combined with the DEV method should reproduce the "observed" (experimental)
DEV spectra, and the input potentials should be tuned until this goal is reached. The situation is familiar from the history of NN
potentials, when many of the subtle details of the NN interactions were fixed from three-body experiments and calculations.

\begin{figure}[h]
%\centering
\includegraphics[width=0.9\textwidth]{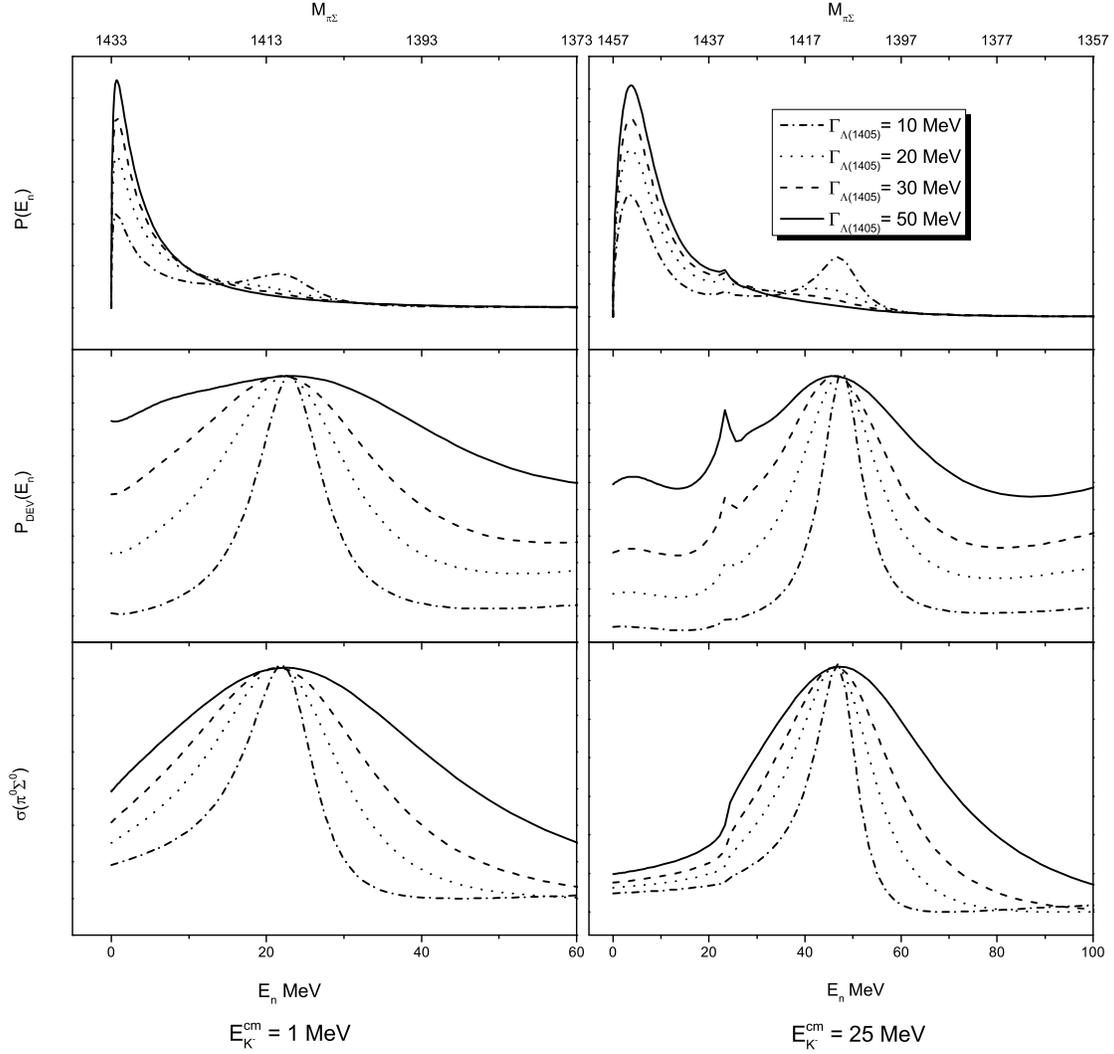}
\caption{Effect of decreasing $\Gamma_{\Lambda(1405)}$ on the neutron spectra
\label{gamma}}
\end{figure}

Finally, we asked the question, under which conditions could the resonance be observed in the direct, $P(E_n)$ spectra. For this purpose we modified two of the $\bar KN-\pi\Sigma$ interaction parameters of one of our potentials (KEK~1), $\lambda^{I=0}_{\bar KN,\bar KN}$ and $\lambda^{I=0}_{\pi\Sigma,\bar KN}$, in such a way, that the position of the $\Lambda(1405)$ remained at its original place, while its width could be made smaller. The results are shown in Fig.\ref{gamma}.

 It is seen, that in order to show up in the direct $P(E_n)$ spectra the width of the original resonance should not exceed $10-15\ MeV$, while the deviation spectra reproduce the original resonance shape in an acceptable way. Thus the real $\Lambda(1405)$ peak, which in all models has a width of $\sim 50-100\ MeV$ has little chance to be seen directly in this reaction, at least in the considered energy region.

%%%%%%%%%%%%%%%%%%%%%%%%%%%%%%%%%%%%%%%%%%%%%%%%%%%%%%%%%%%%%%%%
%%%%%%%%%%%%%%%%%%%%%%%%%%%%%%%%%%%%%%%%%%%%%%%%%%%%%%%%%%%%%%%%%%

\section{Conclusions}
Dynamically exact Faddeev-type calculations for the $K^-+d\rightarrow \pi+\Sigma + n$ reaction were performed in the energy range $E^{cm}_{K^-}=0-50\ MeV$ in order to find the signature of the $\Lambda(1405)$ resonance in the observable neutron spectra.
Four different phenomenological $\bar KN-\pi\Sigma$ interactions were used, all well reproducing the experimental data in the two-body sector, but having rather different pole structure. It was shown, that due to strong kinematical masking effect the inclusive neutron spectra do not exhibit a peak, corresponding to the $\Lambda(1405)$ resonance and show no difference for the four potentials. These spectra are in agreement with the only available experimental data \cite{Tan}. We demonstrated, that the deviation spectrum method is able to eliminate the disturbing kinematical factors and differentiate between potentials with different pole structure. In most of the cases the "deviation" spectra show maxima, which can be related to the "original" two-body $\Lambda(1405)$ resonance. However, the shape and position of the peaks in the deviation spectra may significantly differ from those of their  "original" counterparts. For one of the potentials (SIDDHARTA 1) even the deviation spectrum does not exhibit a maximum, probably due to the closeness of the original pole to the $\bar{K}N$ threshold and its large imaginary part.

\begin{acknowledgments}
The work was supported by the OTKA grant T71989.
\end{acknowledgments}

\end{document}